\documentclass[english,prl,showpacs,reprint,superscriptaddress]{revtex4-2}
\usepackage[latin9]{inputenc}
\usepackage{color}
\usepackage{babel, array, multirow}
\usepackage{verbatim}
\usepackage{bm}
\usepackage{amsmath}
\usepackage{amssymb}
\usepackage{graphicx}
\usepackage{gensymb}
\usepackage[unicode=true,pdfusetitle,bookmarks=true,bookmarksnumbered=false,bookmarksopen=false,breaklinks=false,pdfborder={0 0 0},pdfborderstyle={},backref=false,colorlinks=true]{hyperref}
\hypersetup{citecolor=blue}
\makeatletter
\usepackage{epstopdf}
\usepackage{amsfonts}
\usepackage{mathrsfs}\usepackage{bm}\usepackage{appendix}\usepackage{color}
\usepackage{txfonts}
\usepackage{float}
\usepackage{epstopdf}\usepackage{ulem}
\usepackage{textcomp}

\newcommand{\Rmnum}[1]{\expandafter\@slowromancap\romannumeral #1@}
\usepackage{CJK}
\usepackage{colortbl}
\makeatother

\begin{document}

\title{Maximal Axion Optical Chirality Enabled by Degenerate Quasi-Bound States in the Continuum
}

\author{Chang-Yin Ji}
\affiliation{State Key Laboratory of Chips and Systems for Advanced Light Field Display, Key Lab of advanced optoelectronic quantum architecture and measurement (Ministry of Education) and School of Physics, Beijing Institute of Technology, Beijing 100081, China}

\author{Chong Wang}
\affiliation{State Key Laboratory of Chips and Systems for Advanced Light Field Display, Key Lab of advanced optoelectronic quantum architecture and measurement (Ministry of Education) and School of Physics, Beijing Institute of Technology, Beijing 100081, China}

\author{Jiafang Li}
\email{jiafangli@bit.edu.cn}
\affiliation{State Key Laboratory of Chips and Systems for Advanced Light Field Display, Key Lab of advanced optoelectronic quantum architecture and measurement (Ministry of Education) and School of Physics, Beijing Institute of Technology, Beijing 100081, China}
\affiliation{School of Optics and Photonics, Beijing Institute of Technology, Beijing 10008l, China} 

\author{Yugui Yao}
\email{ygyao@bit.edu.cn}
\affiliation{State Key Laboratory of Chips and Systems for Advanced Light Field Display, Key Lab of advanced optoelectronic quantum architecture and measurement (Ministry of Education) and School of Physics, Beijing Institute of Technology, Beijing 100081, China}

\thanks{Chang-Yin Ji and Chong Wang contributed equally to this work.}

\begin{abstract}
Axion quasiparticles introduce magnetoelectric coupling in topological materials, yet their optical signatures are intrinsically weak, severely limiting experimental detection and practical exploitation. Here, we demonstrate maximal optical chirality of axion quasiparticles by coupling them with degenerate quasi-bound states in the continuum (qBICs). By integrating three-dimensional topological insulators (3DTIs) with achiral photonic crystal slabs (PCSs), we reveal that axion quasiparticles induce an effective Zeeman effect that lifts the degeneracy of achiral qBICs in PCSs, splitting them into two non-degenerate high-$Q$ chiral qBICs with maximally opposite chirality. These result in a more than  $10^3$- and 200-fold increase in polarization conversion efficiency and Faraday rotation angle, respectively, compared to the intrinsic chiroptical response of bare 3DTI films. Moreover, circular dichroism and nonreciprocal optical responses are revealed as new key signatures for probing axion quasiparticles in 3DTIs-PCSs coupled systems. Our results provide a definitive route for observing axion quasiparticles and topological physics, while establishing a promising platform for controlling optical polarization, spin, phase, and nonreciprocal responses.
\end{abstract}

\maketitle
Axion are hypothetical elementary particles originally proposed to resolve the strong charge-parity problem in quantum chromodynamics and are now considered promising candidates for dark matter \cite{peccei1977cp, weinberg1978new, sikivie1983experimental,wilczek1987two, chadha2022axion, duan2023resonant}. Although axions have not yet been observed in high-energy physics, recent advances in condensed-matter physics have revealed that axion can emerge in solid-state systems as collective quasiparticles, particularly in topological materials such as three-dimensional topological insulators (3DTIs), quantum anomalous Hall systems, axion insulators and Weyl semimetals \cite{qi2008topological,tse2010giant,maciejko2010topological,tse2010magneto,jenkins2010terahertz,swingle2011correlated,valdes2012terahertz,grushin2012finite,shuvaev2013terahertz,baasanjav2014magnetoelectric,wu2015high,morimoto2015topological,wu2016quantized,okada2016terahertz,dziom2017observation,mogi2017magnetic,zhang2019topological,tokura2019magnetic,xu2019higher,liebman2025multiphoton}. In these systems, axions generate magnetoelectric responses, in which a quantized or dynamical axion field $\theta$, determined by the Berry-curvature topology of the electronic band structure, couples electric and magnetic polarization through a universal $\bm E\cdot \bm B$ term  \cite{qi2008topological,sekine2021axion}. Because $\theta$ is a pseudoscalar, this coupling endows light-matter interactions with intrinsic chirality. Axion optical chirality manifests itself as Faraday and Kerr rotation, providing a direct optical probe of axion physics and fundamental physical constants in condensed matter \cite{maciejko2010topological, wu2016quantized}. However, the optical chirality of axions is intrinsically weak, typically producing minute polarization rotation angles and negligible linear polarization-conversion efficiencies \cite{tse2010giant,maciejko2010topological,tse2010magneto,wu2016quantized, ahn2022theory,shaposhnikov2023emergent, seidov2025dual,shaposhnikov2025probing,jazi2025realization}. Until now, a viable strategy to significantly boost the optical chirality of axions remains elusive in these fields, severely constraining both experimental detection and technological exploitation.

A natural approach to enhance chiroptical responses is to exploit  resonant light-matter interaction with high-quality (high-$Q$) factors. Quasi-bound states in the continuum (qBICs) represent an unconventional paradigm for
realizing high-$Q$ optical resonances that strongly enhance light-matter interactions \cite{neumann1929merkwurdige,marinica2008bound,hsu2016bound, kang2023applications}. However, qBICs supported by mirror-symmetric structures are intrinsically achiral and thus incapable of asymmetric coupling to the spin angular momentum of photons \cite{koshelev2018asymmetric,li2019symmetry, han2019all,abujetas2021high,aigner2025optical}. 
Recently, chiral qBICs have been realized by intentionally breaking geometric mirror symmetries, providing a promising approach
to achieve maximum chirality \cite{gorkunov2020metasurfaces,overvig2021chiral,gorkunov2021bound, shi2022planar, zhang2022chiral, chen2023observation,kuhner2023unlocking,gromyko2024unidirectional, gromyko2025enabling}, resulting in
chiral lasing and giant circular dichroism. However, these approaches are constrained by the fact that increasing geometric chirality not only complicates device design but also inevitably lowers the $Q$ factors of chiral qBICs. Furthermore, the reported chiral qBICs lack robustness due to their extreme sensitivity to geometric parameters, making them highly vulnerable to structural perturbations. 

Here, we demonstrate a strategy to realize robust high-$Q$ maximal axion optical chirality by coupling axion quasiparticles in 3DTIs with degenerate qBICs in achiral photonic crystal slabs (PCSs). Our theory reveals that the axion quasiparticles generate an effective Zeeman field that converts degenerate achiral qBICs into two non-degenerate high-$Q$ chiral qBICs with completely opposite maximum chirality, which significantly enhances the optical signatures of the weak axion field, including polarization conversion efficiency, Faraday rotation angle, circular dichroism and nonreciprocal optical responses. This work not only provides a new pathway to enhance the coupling between axions and photons, but also paves the way for controlling optical polarization, spin, phase and nonreciprocal responses.

\begin{figure*}\includegraphics[width=0.95\textwidth]{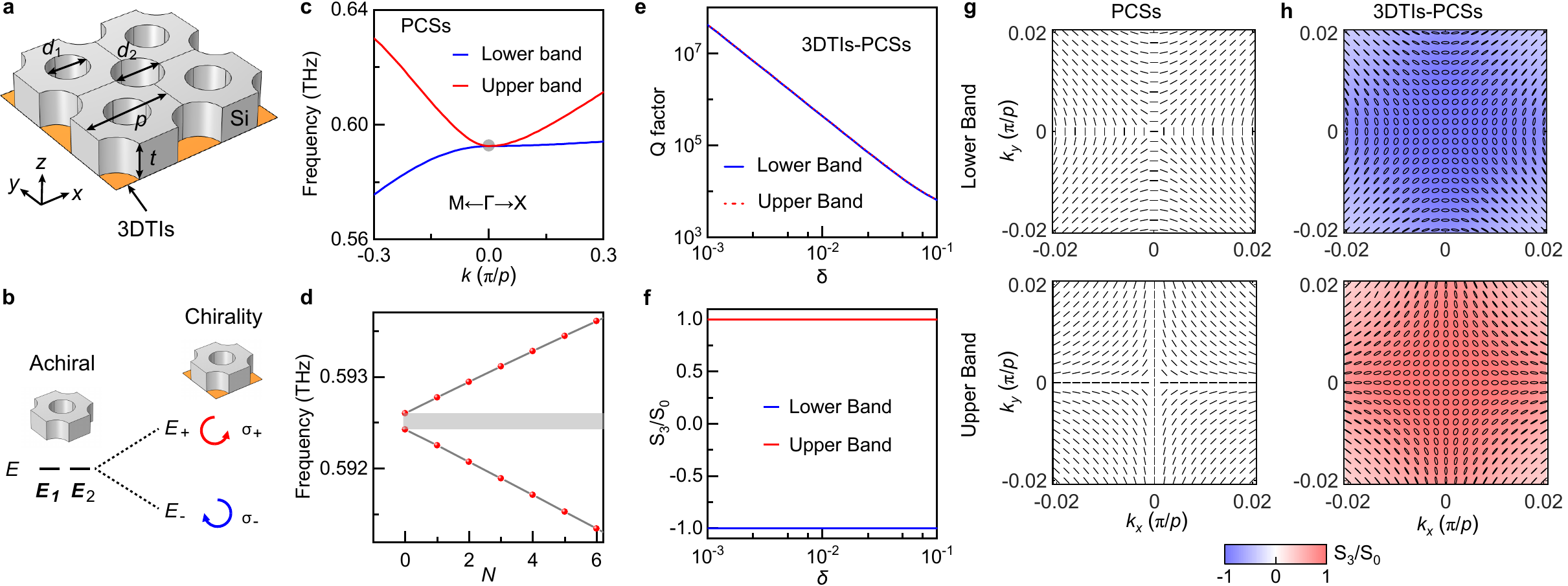}
\caption{Maximal axion optical chirality. (a) Schematic illustration of the proposed 3DTIs-PCSs coupled system with a square lattice. The lattice period and thickness of the PCSs are denoted as $p$ and $t$, respectively. The diameters of the central and corner cylindrical air holes are represented by $d_1=(0.5-\delta)p$ and $d_2=(0.5+\delta)p$. (b) Schematic illustration of the spin-locked splitting induced by axion quasiparticles, where the degenerate achiral qBICs ($\bm E_1$ and $\bm E_2$) in achiral PCSs are lifted in the 3DTIs-PCSs coupled system, giving rise to two high-$Q$ chiral qBICs ($E_+$ and $E_-$) with completely opposite chirality ($\sigma_+$ and $\sigma_-$). (c) Calculated band structure of the PCSs for parameters $p=200$ $\mu$m, $t=0.75p$ and $\delta=0.04$. The gray circle indicates the position of the doubly degenerate achiral qBICs in PCSs. (d) Zeeman splitting versus highest fully occupied Landau level $N$ for $\delta=0.04$. (e) Dependence of the $Q$ factor of the chiral qBICs on the $\delta$ for $N=0$. (f) Normalized third Stokes parameter $S_3 /S_0$ versus $\delta$ for $N=0$. (g, h) Momentum-space distribution of polarization states for the PCSs at $\delta=0.04$ and for the 3DTIs-PCSs coupled systems at $N=0$ and $\delta=0.04$. The refractive index of silicon is taken as $n_{Si}=\sqrt{11.7}$, while the surrounding medium has a refractive index of $n_{air}=1$.}
\label{Fig-1}
\end{figure*}

We consider the PCSs preserve $C_{4v}$ symmetry as a general and illustrative example, with the gray-colored bulk geometry shown in Fig.~\ref{Fig-1}(a). Here, the PCSs lack geometric chirality due to they retain all mirror symmetries. The yellow region represents the 3DTIs. Axion electrodynamics of 3DTIs modifies Gauss's and Amp$\grave{e}$re's laws as follows,
\begin{equation}
\begin{aligned}
  \nabla \cdot \bm E&=-2c\alpha \nabla (\frac{\theta}{2\pi}) \cdot \bm B\\
  \nabla \times\bm B &=\frac{1}{c^2}\frac{\partial \bm E}{\partial t}+\frac{2\alpha}{c}[\frac{\partial}{\partial t}(\frac{\theta}{2\pi})\bm B+\nabla (\frac{\theta}{2\pi})\times\bm E]
\end{aligned}
\label{L_0}
\end{equation}
Here, we consider source-free cases. $\alpha$ is the fine-structure constant, $c$ is the speed of light and $\theta$ is the topological contribution. With $\theta=\pi$~(mod $2\pi$) being a time- and space-independent constant in 3DTIs, the topological contribution does not alter the bulk wave equation but instead appears solely as a surface current $\bm J_{s}=-\theta\alpha /\pi Z_0\hat{n}\times\bm E_t$ at the surface of 3DTIs, where $\hat{n}$ is the normal vector of the boundary, $\bm E_t$ is the tangential component of the electric field at the boundary, $Z_0$ is the characteristic impedance of vacuum (see Sec.1 in \cite{jiprl}). The optical response of 3DTIs can be entirely characterized by $\bm J_{s}$, as the longitudinal conductivity vanishes in the quantum regime and non-topological current is absent when the chemical potential is located within the band gap \cite{wu2016quantized}. Furthermore, the thickness of 3DTIs is far smaller than the wavelength of light at terahertz frequencies. Thus, the contributions from the two conducting surfaces of 3DTIs can be modeled as a single surface with $\bm J_{s}=-2\alpha (N_t+N_b+1)/Z_0\hat{n}\times\bm E_t$ \cite{wu2016quantized}. $N_t$ and $N_b$ represent the highest fully occupied Landau level (LL) of the 3DTI film's top and bottom surfaces, respectively.

We next theoretically analyze the coupling between the axion field in 3DTIs and photonic modes of the PCSs. The $\Gamma$ point correspondingly supports a two-dimensional irreducible representation of $E$ with two degenerate orthogonal modes ($\bm E_1$ and $\bm E_2$) for PCSs with $C_4$ rotational symmetry \cite{sakoda2005optical}, as shown in Fig.~\ref{Fig-1}(b). The interaction between the axion and the two degenerate orthogonal modes supported by the PCS is described by the following Hamiltonian $\cal{H}$ (see Sec.2 in \cite{jiprl}),
\begin{equation}\label{ham}
  {\cal H}=\left(
                              \begin{array}{cc}
                                \omega {_0}-i\gamma & -i\kappa \\
    i\kappa & \omega {_0}-i\gamma \\
                              \end{array}
                            \right)
\end{equation}
where $\omega_0$ and  $\gamma$ denote the angular frequency and decay loss of the degenerate optical modes in the PCSs, respectively.  $\kappa= -\theta\alpha c/2\pi\int \bm E_1\times \bm E_2^* \cdot d \bm s$ represents the coupling strength between $\bm E_1$ and $\bm E_2$ induced by the $\bm J_s$. The complex eigenfrequencies of $\cal{H}$ are as follow,
\begin{equation}
\omega{_\pm}=\omega{_0}-i\gamma \pm \kappa , \label{eng}
\end{equation}

For $\theta= 0$ (i.e., without axion quasiparticles), the coupling $\kappa$ vanishes, resulting in a pair of frequency degenerate linearly polarized eigenstates of the $\cal{H}$: $\bm E_1=[1,0]^T$ and $\bm E_2=[0,1]^T$. The presence of axion quasiparticles makes $\kappa \neq 0$, effectively introducing a Zeeman-like field that lifts the frequency degeneracy of the optical modes, resulting in $\omega_- \neq \omega_+$. Another interesting phenomenon is that for $\kappa \neq 0$, the two eigenstates of $\cal{H}$ are circular eigenstate,
\begin{equation}
\sigma_\pm =[1,\pm i]^T
\end{equation}
 In such a case, we can obtain maximized axion optical chirality, as shown in Fig.~\ref{Fig-1}(b). Here, the maximal chirality corresponds to purely circularly polarized far-field radiation from the resonant modes. Moreover, we can see that the formation of the circularly eigenstates is independent of the PCSs parameters, and therefore the maximized axion optical chirality is expected to be highly robust against perturbations to the PCSs. If, in addition, an extremely small $\gamma$ can be achieved, robust high-$Q$ maximized axion optical chirality can be realized.

\begin{figure}\includegraphics[width=0.47\textwidth]{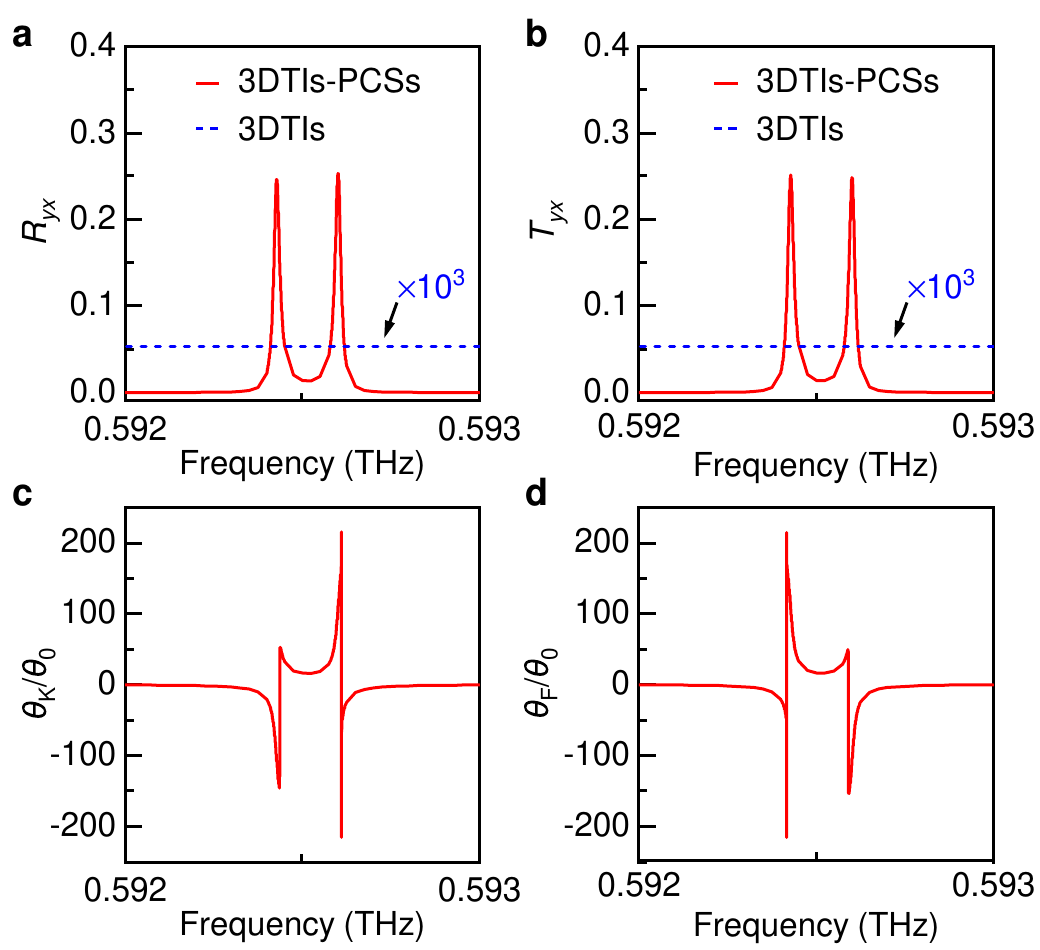}
\caption{Giant polarization rotation induced by high-$Q$ maximal axion optical chirality. (a, b) Simulated reflection ($R_{yx}$) and transmission ($T_{yx}$) spectra for linearly polarized cross-polarization component. Here, $R_{yx}=|r_{yx}|^2$, $T_{yx}=|t_{yx}|^2$. Owing to the $C_4$ rotational symmetry, it follows that $|r_{yx}|^2=|r_{xy}|^2$ and $|t_{yx}|^2=|t_{xy}|^2$, see Sec.3 in the \cite{jiprl}. For clarity, the results for the 3DTIs in (a, b) are multiplied by $10^3$. (c, d) Calculated Kerr and Faraday rotation angles ($\theta_K$ and $\theta_F$). $\theta_0=\arctan(1/137)$. The incident direction of linearly polarized light is along the $+z$ direction.}
\label{Fig-2}
\end{figure}

Next, we present a specific example to elucidate such intriguing phenomenon. The PCSs have a period $p=200$ $\mu$m and thickness $t=0.75p$, with each unit cell incorporating two cylindrical air holes of diameters $d_1=(0.5-\delta)p$ and $d_2=(0.5+\delta)p$, as shown in Fig.~\ref{Fig-1}(a). All numerical simulations are performed using the commercial software COMSOL Multiphysics (see Sec.4 in the \cite{jiprl}). Fig.~\ref{Fig-1}(c) shows the calculated band structure of the PCSs for $\delta=0.04$. We have doubly degenerate modes at a frequency of 0.59252 THz at the $\Gamma$ point, marked as gray circle in Fig.~\ref{Fig-1}(c).

Fig.~\ref{Fig-1}(d) illustrates the frequency evolution of the at-$\Gamma$ degenerate modes in Fig.~\ref{Fig-1}(c) in the presence of axion field for $\delta=0.04$. Here, we have $2N=N_t+N_b$. It can be seen that the axion effect indeed induces frequency splitting, which exhibits a linear dependence on $N$, in agreement with the prediction of Eq.~\ref{eng}. The frequency splitting in the quantum limit ($N=0$) is highlighted by the gray shaded region in Fig.~\ref{Fig-1}(d), which can be attributed to the half-integer quantum Hall effect on the surface of 3DTIs \cite{hasan2010berry}. Fig.~\ref{Fig-1}(e) shows the variation of the $Q$ factor as a function of $\delta$ for 3DTIs-PCSs at $N=0$, exhibiting the trend $Q\propto \delta ^{-2}$ for small $\delta$. In such a case, BICs with an infinite $Q$ factor can be obtained at $\delta = 0$. For $\delta \neq 0$, we can obtain high-$Q$ qBICs.

We use the normalized third Stokes parameter $S_3 /S_0$ to quantifies the degree of chirality of the axion quasiparticles \cite{ji2024probing,ji2025intrinsically}, as shown in Fig.~\ref{Fig-1}(f). It can be seen that the values of $S_3 /S_0$ approach -1 and 1 for the lower and upper band qBICs, respectively. This confirms that we have realized maximized axion optical chirality even in the quantum limit ($N=0$) of 3DTIs in 3DTIs-PCSs coupled system. Here, the emergence of two circular eigenmodes with opposite handedness and near-unity $S_3 /S_0$ at the split resonances is intrinsically rooted in the physics of the degenerate qBICs, as shown in Fig.~\ref{Fig-1}(d, f). The chiral resonances can simultaneously exhibit maximal chirality and high-$Q$ factors, as shown in Fig.~\ref{Fig-1}(e, f)]. This is due to we can independently harness the axion effect to induce maximal optical chirality and qBICs to tune the $Q$ factor, rather than relying on breaking the mirror symmetry of the system through bulk geometry. More interestingly, since the value of $S_3/S_0$ is independent of $\delta$ in Fig.~\ref{Fig-1}(f), indeed enabling robust high-$Q$ maximized axion optical chirality. However, chiral resonant modes arising from bulk geometric chirality strongly depend on the geometric parameters and fragile to perturbation of geometric parameters \cite{gorkunov2020metasurfaces,overvig2021chiral,gorkunov2021bound, shi2022planar, zhang2022chiral, chen2023observation,kuhner2023unlocking,gromyko2024unidirectional, gromyko2025enabling}.

The axion-induced chiral polarization states in momentum space are further explored. For comparison, the polarization states in PCSs are shown in Fig.~\ref{Fig-1}(g). The polarization states of both the upper and lower bands are seen to exhibit linear polarization across momentum space. Thus, the achiral PCS itself exhibits no optical chirality. However, in the 3DTIs-PCSs coupled system, the polarization states of both the lower and upper bands become nearly circular near the $\Gamma$ point due to the axion effect in 3DTIs, as shown in Fig.~\ref{Fig-1}(h). Consequently, the 3DTIs-PCSs coupled systems exhibit chirality even the PCSs are achiral, which is in sharp contrast
to prior approaches that achieve chirality through chiral geometric features \cite{kong2020plasmonic,mun2020electromagnetic,crassous2023materials,lininger2023chirality,dong2024emerging,wang2024chiral,yan2024structural,deng2024advances}. In addition, the fabrication of achiral PCSs is considerably simpler than that of chiral PCSs and metasurfaces.

\begin{figure*}\includegraphics[width=0.8\textwidth]{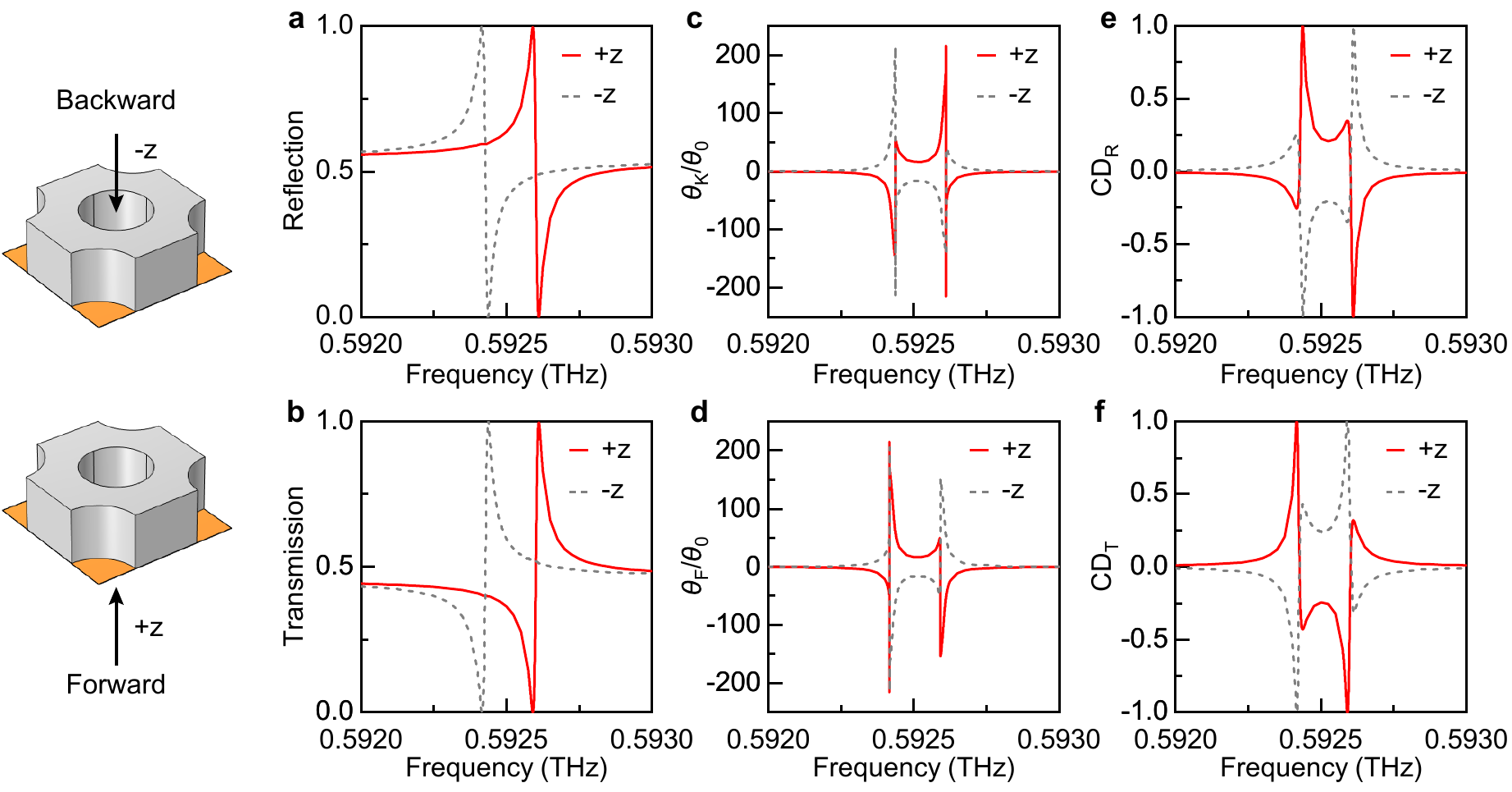}
\caption{Pronounced nonreciprocity and circular dichroism enabled by high-$Q$ maximal axion optical chirality. (a) Reflection spectra of cross-polarized circular components $R_{LR}=|r_{LR}|^2$ for incident RCP light propagating along the $+z$ and $-z$ (forward and backward) directions. (b) Transmission spectra of co-polarized circular components $T_{RR}=|t_{RR}|^2$ for incident RCP light propagating along the $+z$ and $-z$ directions. (c, d) Kerr and Faraday rotation angles for excitation light propagating along the $+z$ and $-z$ directions. (e, f) Reflection and transmission circular dichroism ($CD_R$ and $CD_T$) spectra for incident circularly polarized light propagating along the $+z$ and $-z$ directions.}
\label{Fig-3}
\end{figure*}

We further demonstrate that realized high-$Q$ maximal axion optical chirality can substantially enhance the optical signatures of axion effect in the quantum limit ($N=0$). Fig.~\ref{Fig-2}(a) and Fig.~\ref{Fig-2}(b) are the cross-polarization reflection ($R_{yx}=|r_{yx}|^2$) and transmission ($T_{yx}=|t_{yx}|^2$) spectra, respectively. Here, $t_{ij}$ ($r_{ij}$) denotes the $i$-polarized transmittance (reflectance) coefficient for $j$-polarized incident light, see Sec.3 in the \cite{jiprl}. It can be seen that the linear polarization conversion efficiency for the reflected and transmitted waves reaches a maximum value of approximately 0.25 near $\omega_-$ and $\omega_+$ in 3DTIs-PCSs coupled system, as shown in Fig.~\ref{Fig-2}(a, b). However, the linear polarization conversion efficiency of reflected and transmitted light is $\alpha ^2/(1+\alpha ^2)^2\sim 5.3245 \times 10^{-5}$ for free-standing 3DTIs in the quantum limit, see the blue dashed line in Fig.~\ref{Fig-2}(a, b) and Sec.5 in the \cite{jiprl}. Thus, the linear polarization conversion efficiency is enhanced by more than $10^3$ times through the maximal axion optical chirality. 

We can also expect to obtain tailored linear polarization rotation effects, quantified by Faraday and Kerr angles ($\theta_F$ and $\theta_K$), 
\begin{equation}
  \theta_F=\frac{arg (t_{LL})-arg (t_{RR})}{2}, \quad \theta_K=\frac{arg (r_{RL})-arg (r_{LR})}{2}
\end{equation}
$t_{RR}$ and $t_{LL}$ ($r_{LR}$ and $r_{RL}$) denote the complex transmission (reflection) coefficients of the co-polarized (cross-polarized) circular components under normally incident left- and right-handed circularly polarized (LCP and RCP) light, respectively. Fig.~\ref{Fig-2}(c) and Fig.~\ref{Fig-2}(d) are the results of  $\theta_K$ and $\theta_F$, respectively. The maximum value of $|\theta_F|$ can be close to $\pi/2$ in the 3DTIs-PCSs coupled systems. However, the $\theta_F$ of the free-standing 3DTIs in the quantum limit is $\theta_F=\arctan\alpha\sim 1/137$, see Sec.5 in the \cite{jiprl}. Thus, it can be observed that the largest $|\theta_F|$ is enhanced by more than 200 times and is significantly larger than that in previous studies \cite{tse2010giant,maciejko2010topological,tse2010magneto,jenkins2010terahertz,valdes2012terahertz,wu2016quantized,okada2016terahertz,ahn2022theory}. It is worth noting that the observed enhancement of the $\theta_F$ arises from the strong enhancement of axion chirality in 3DTIs-PCSs coupled systems, together with the intrinsically weak Faraday response of bare 3DTI films [see Fig.~\ref{Fig-2}(d) and Fig.~S3(g) in the \cite{jiprl}]. In contrast, the $\theta_K$ does not exhibit enhancement, since the $\theta_K$ of bare 3DTI films is already maximized [see Fig.~\ref{Fig-2}(c) and Fig.~S3(h) in the \cite{jiprl}]. Although no amplification of $\theta_K$ is observed in 3DTIs-PCSs coupled systems, the maximum reflected light signal is markedly enhanced more than $10^3$ times, facilitating the experimental robust detection of the Kerr angle.

Finally, in addition to linear polarization rotation and conversion, we uncover that nonreciprocity and circular dichroism can act as new indicators for detecting the axion quasiparticles based on the high-$Q$ maximal axion optical chirality. Nonreciprocity refers to the fact that interchanging the positions of the source and the observer results in different electromagnetic responses. Thus, the observed electromagnetic response depends on the propagation direction of light in a nonreciprocal optical system. Although the axion quasiparticles naturally give rise to nonreciprocal optical behavior, this phenomenon has not yet been observed in 3DTIs. Here, the nonreciprocal optical responses are revealed through the reflection, transmission, $\theta_F$, $\theta_K$, as well as circular dichroism spectra. We interchange the positions of the source and the observer by changing the propagation direction of the normally incident light. Specifically, the reflection ($R_{LR}=|r_{LR}|^2$) in Fig.~\ref{Fig-3}(a) and the transmission ($T_{RR}=|t_{RR}|^2$) spectra in Fig.~\ref{Fig-3}(b) indicate that forward-propagating RCP light couples exclusively to the high-frequency resonance, whereas backward-propagating RCP light couples only to the low-frequency resonance, confirming the nonreciprocal optical behavior. Furthermore, upon reversing the propagation direction, both the Kerr and Faraday angles reverse in sign, offering further confirmation of the system's nonreciprocal behavior, as illustrated in Fig.~\ref{Fig-3}(c, d).

The circular dichroism has been well explored in chiral materials \cite{liu2018nano,liu2020fano,ji2021artificial,crassous2023materials}, while it arises from the axion quasiparticles in 3DTIs remains largely unexplored. We define the reflection and transmission circular dichroism ($CD_R$ and $CD_T$) as follow,
\begin{equation}
 CD_R=\frac{|r_{LR}|^2-|r_{RL}|^2}{|r_{LR}|^2+|r_{RL}|^2}, \quad CD_T=\frac{|t_{RR}|^2-|t_{LL}|^2}{|t_{RR}|^2+|t_{LL}|^2}
\end{equation}
As shown in Fig.~\ref{Fig-3}(e, f), we can get both $CD_R$ and $CD_T$ reach to $\pm1$ near the resonance due to the maximal axion optical chirality in 3DTIs-PCSs coupled systems, realizing maximum circular dichroism response of axion quasiparticles. In such case, this phenomenon can be exploited to achieve nearly perfect conversion from linear to circular polarization devices using 3DTIs. The $CD$ spectra of both the reflected and transmitted waves change when the propagation direction of the normally incident light is reversed, reflecting the system's nonreciprocal response, as shown in Fig.~\ref{Fig-3}(e, f). Chiral structures can also give rise to direction-dependent optical responses in reciprocal systems \cite{pan2021asymmetrical}. Therefore, in determining nonreciprocity via the scattering parameters $t_{ij}^{\alpha\beta}$ and $r_{ij}^{\alpha\alpha}$ (see Sec. 3 in \cite{jiprl}), it is essential to consider both the interchanging the positions of the source and the observer and the polarization states of the incident and outgoing waves, i.e. $t_{ij}^{\alpha\beta}\neq t_{ji}^{\beta\alpha}$ and $r_{ij}^{\alpha\alpha}\neq r_{ji}^{\alpha\alpha}$ (see Eq.~28 in Sec.3 of the \cite{jiprl}).

Related chiral and nonreciprocal effects have also been explored in magneto-optical photonic systems and topological photonic structures~\cite{zhao2024spin,ye2025chiral,lai2025photonic}, where they are typically associated with bulk magneto-optical responses or higher-order topological transport (see Sec.7 in the \cite{jiprl}). In contrast, our work leverages the coupling between axionic surface responses in topological electronic materials and degenerate qBICs in achiral PCSs, yielding resonantly enhanced far-field chiroptical signatures for probing axion quasiparticles and topological phenomena in condensed-matter systems, while also providing a versatile platform for controlling light-matter interactions beyond conventional magneto-optical materials.

We propose a feasible experimental scheme to realize and observe the enhanced axion chirality. 3DTI films, such as Bi$_2$Se$_3$, can be epitaxially grown on Si(111) using molecular beam epitaxy~\cite{he2011epitaxial,bansal2011epitaxial,wang2011growth,gao2017patterning}. Then, the systems can be readily fabricated using deep reactive ion etching or inductively coupled plasma etching. To protect the 3DTI film, the silicon can be designed to be partially etched rather than fully etched. Optical characterization can be carried out using mature terahertz time-domain spectroscopy~\cite{wu2016quantized}. For Bi$_2$Se$_3$, the reported response can be observed under a magnetic field of $B\geq 5$~T and temperature $T\leq 5$~K~\cite{wu2016quantized}. Si loss and longitudinal conductivity can affect the $Q$ factor, linear polarization conversion efficiency and the enhancement of the Faraday rotation, while having negligible impact on mode splitting, nonreciprocal optical responses and $S_3/S_0$ (see Sec.8-9 in the \cite{jiprl}). Therefore, employing high-resistivity float-zone silicon~\cite{dai2004terahertz}, a uniform 3DTI film, and low-temperature operation can effectively mitigate the impact of material losses on these optical effects. The impact of small fabrication imperfections on the axion optical chirality in the 3DTIs-PCSs coupled system is negligible (see Sec.10 in the \cite{jiprl}). $CD$ at split resonances, forward/backward spectral contrast, Faraday signatures and momentum-space chirality can be used to verify the predicted axion-induced chiral eigenmodes.

In summary,  we propose a novel scheme to realize robust maximal axion optical chirality through the interplay between the axion quasiparticles in 3DTIs and the degenerate qBICs in achiral PCSs, which is uncovered by the established theory. We further show that the realization of maximal axion optical chirality in 3DTIs-PCSs coupled system not only substantially boosts the polarization conversion efficiency and Faraday rotation angle associated with weak axion fields, but also generates pronounced circular dichroism and optical nonreciprocity, which offer new optical signatures for probing axion fields in condensed matter systems. Furthermore, our study can also provide new perspectives for enhancing axion responses in other topological systems, such as magnetic Weyl semimetals. Finally, our findings also establish a new strategy for realizing high-$Q$ chiral resonances and for developing advanced chiral and nonreciprocal optoelectronic devices enabled by topological materials.

{\textbf{Acknowledgments:}  Chang-Yin Ji and Chong Wang contributed equally to this work. The authors acknowledge the
support from the National Natural Science Foundation of
China (Grants No.W2511003,No. 12321004, No. 12204041,
No. 12274030,No.T2325005,No. 52488301,No. 12574335,
No. 12234003, No. 62375016), and National Key
Research and Development Program of China (Grants
No. 2022YFA1403400 and No. 2024YFB2809204), and
the Fundamental and Interdisciplinary Disciplines
Breakthrough Plan of the Ministry of Education of China
(Grant No. JYB2025XDXM115).


\end{document}